\documentclass[twocolumn,aps,prb,epsf,showpacs,superscriptaddress]{revtex4}
\usepackage{amssymb}
\usepackage{amsmath}
\usepackage{epsfig}
\usepackage{epstopdf}
\usepackage{bbm}
\usepackage{graphicx}
\usepackage{color}

\begin{document}
\bibliographystyle{apsrev}

\title{Chemical Control of Orbital Polarization in Artificially Structured Transition Metal Oxide Materials: the case of $La_2NiXO_6$}

\author{M. J. Han}
\affiliation{Department of Physics, Columbia University, New York, New
  York 10027, USA}

\author{C. A.  Marianetti }
\affiliation{Department of Applied Physics,Columbia University, New
  York, New York 10027, USA }

\author{A. J. Millis }
\affiliation{Department of Physics, Columbia University, New York, New
  York 10027, USA}

\date{\today }

\begin{abstract}
The application of modern layer-by-layer growth techniques to transition metal oxide materials raises the possibility of creating new classes of materials with rationally designed correlated electron properties. An important step towards this goal is the demonstration that electronic structure can be controlled by atomic composition. In compounds with partially occupied transition metal $d$-shells, one important aspect of the electronic structure is the relative occupancy of different $d$-orbitals . Previous work has established that strain and quantum confinement can be used to influence orbital occupancy. In this paper we demonstrate a different modality for orbital control in transition metal oxide heterostructures, using density functional band calculations supplemented by a tight binding analysis to show that the choice of non-transition-metal counterion $X$ in transition-metal oxide heterostructures composed of alternating $LaNiO_3$ and $LaXO_3$ units strongly affects orbital occupancy, changing the magnitude and in some cases the sign of the orbital polarization.
\end{abstract}


\maketitle

\section{Introduction}

Recent progress in growth of  transition metal oxide multilayers with atomic-scale chemical precision \cite{Mannhart08} suggests that it may become possible to create new classes of materials  with desirable electronic properties based on aspects of correlated electron physics such as high Curie temperature ferromagnetism\cite{Kobayashi98,Luders09}, `colossal' magnetoresistance\cite{Jin94}, correlation-driven metal-insulator transitions\cite{Imada98} and high transition temperature superconductivity\cite{Bednorz86}.  The ultimate goal is `materials by design', in other words to construct systems with desired electronic properties. A  necessary first step is to design and fabricate systems with a desired electronic structure.  

The `correlated electron' properties of transition metal oxides are controlled in part by the relative occupancy of the different transition-metal $d$ orbitals.\cite{Nagaosa00} Controlling the orbital occupancy by materials design is therefore an important milestone in the progress towards a  rational design of correlated electron materials. A difference in relative occupancies of orbital states may be described as an `orbital polarization' in analogy to the difference in occupancies of spin states which gives rise to spin polarization. As the control of spin polarization is achieved by application of appropriate magnetic fields, so the control of orbital polarization may be achieved by identification and manipulation of appropriate `orbital fields'. 

Two classes of `orbital field'  are well established: strain and quantum confinement. Lattice strain works because  in a transition metal oxide  the hybridization between the transition metal $d$ orbital and the oxygen $p$ states produces ligand fields whose strength depends on the geometrical distance between the transition metal and oxygen site. An applied strain changes relative bond lengths, thereby affecting the ligand fields.   Strain control was reported by Konishi and collaborators\cite{Konishi99}, who grew films of `colossal magnetoresistance' (CMR) manganites on substrates which imparted compressive, negligible, or tensile strain to the manganite film. The systems exhibited resistivities which were respectively strongly insulating, weakly insulating, and metallic and the change in resistance was attributed to a strain-induced change in relative occupancies of the $d_{x^2-y^2}$ and $d_{3z^2-r^2}$ orbitals on the electrically active $Mn$ site. 

The quantum confinement effect works because each transition metal $d$-orbital has a specific spatial structure which leads to a direction-dependent hopping amplitude. Spatially anisotropic quantum confinement (for example in a heterostructure composed of alternating transition metal oxide and insulating spacer layers) allows electrons to delocalize more in some directions than in others, thereby allowing some orbitals to gain more delocalization energy than others. The quantum confinement effect  operates in a slightly subtle way in transition metal oxide systems, where the relevant electronic bands are antibonding combinations of transition metal $d$ and oxygen $p$ states. In the single crystal form of many transition metal oxide materials the transition metal-oxygen hybridization vanishes at the zone center ($\Gamma$) point of the Brillouin zone. The bands are degenerate or nearly degenerate at this point, and disperse upwards from it. A partial breaking of translational symmetry activates the hopping at zone center, thus lifting some orbitals up relative to others. Chaloupka and Khalliulin \cite{Chaloupka08}  recently proposed that in a  heterostructure composed of alternating layers of $LaNiO_3$ and $LaAlO_3$ this effect could lead to a situation in which only the $x^2-y^2$ $Ni$ $e_g$ symmetry orbital would be occupied, providing an electronic structure similar to that found in high $T_c$ copper-oxide superconductors. The issue was studied theoretically by Hansmann and collaborators \cite{Hansmann09}.  

In this paper we identify  a third route to control of orbital polarization, based on chemical composition of spacer layers in an oxide superlattice.  We consider specifically the class of systems introduced by Chaloupka and Khaliullin \cite{Chaloupka08}, namely superlattices composed of alternating layers of $LaNiO_3$ and of a spacer layer $LaXO_3$ the orbital polarization depends upon the choice of counterion $X$, even when $X$ is chosen such that $LaXO_3$ is a wide-bandgap insulator.   In this situation the conventional wisdom is that because the $LaXO_3$ layers are insulating, the only relevant effects are strains induced by lattice mismatch between the $LaNiO_3$ and the $LaXO_3$ layers.  We show that this is not the case, and that the strength of chemical bonding of the apical oxygen with the $X$ ion significantly affects the polarization, even changing its sign in some cases.   

The rest of this paper is organized as follows. Section \ref{Model} presents the systems we study, the definitions we use and the formalism we apply. Section \ref{Results} presents the band structures and resulting orbital polarizations in the simple case where lattice relaxations are forbidden. Section \ref{tb} presents a tight binding analysis which explicates  the physics behind the results presented in  Section \ref{Results}. Section \ref{Relax} considers the additional effects of lattice relaxation. Section \ref{Conclusion} is a summary and conclusion.

\section{Model, Definitions and Methods\label{Model}}

\begin{figure}[b]
\begin{center}
\includegraphics[width=0.8\columnwidth,angle=0]{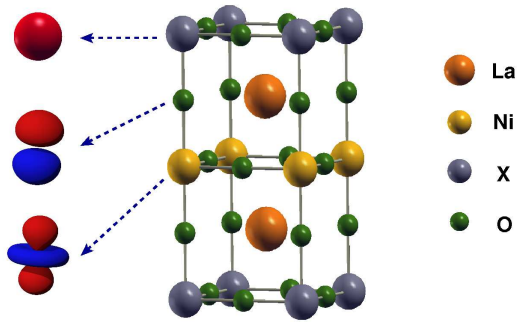}
\caption{ Central portion: Unit cell of $(001)$ superlattice considered in  this paper with atoms indicated on right.  Left side: chain of hybridizing  orbitals (top to bottom $X_s$, $O_{p_z}$ and $Ni_{3z^2-r^2}$)  controlling relative occupancy of $d_{3z^2-r^2}$ orbital.}
\label{geometry}
\end{center}
\end{figure}

We consider superlattices of the form depicted in Fig. \ref{geometry} composed of  blocks of $LaNiO_3$ and another material $LaXO_3$, alternating along the $(001)$ axis of the basic perovskite crystal structure.  We  choose $X$=$B$, $Al$, $Ga$, $In$ from column $3A$ of the periodic table. While to our knowledge only $LaAlO_3$ has been used to grow oxide superlattices, all of the $LaXO_3$ compounds   have been reported in the literature \cite{Ruiz-Trejo03,Fan06,Chaloupka08}.  For simplicity we take all $Ni-O-Ni$ and $Ni-O-X-O-Ni$ bonds angles to be $180^\circ$. In most of our calculations we fix the in-plane lattice constants to the GGA-optimized pseudocubic bulk $LaNiO_3$ value $3.81\AA$ but in  some of our calculations the in-plane lattice constant was set to $3.91\AA$ to model the effect of a film grown on an $SrTiO_3$ substrate.  We consider two cases, an `unrelaxed' case where the out of plane lattice constant is set equal to the in-plane one, and a `relaxed' structure where the out of plane atomic positions are adjusted to minimize the energy as described below.

To determine the electronic structure we use density functional band theory within the Generalized Gradient Approximation (GGA) as implemented in the Vienna ab-initio simulation package (VASP)\cite{Kresse1993558,Kresse199414251,Kresse199615,Kresse199611169}.  We used a plane-wave basis set and the projector augmented wave method\cite{Kresse19991758} with a  cutoff of $270$ $eV$ and $k$-point meshes of $10\times10\times5$.  In the `relaxed' calculations, atomic positions were relaxed along the $(001)$ using conjugate gradient minimization of the GGA energy. 

To determine orbital occupancies we projected the calculated electronic density of states onto locally defined atomic orbitals obtained by defining a sphere around the atom in question and then projecting the wave functions within the sphere onto the appropriate symmetry states. The sphere sizes in $\AA$ were taken to be $1.286$ ($Ni$-$d$) $0.820$ ($O$-$p$) (these are the VASP defaults) while for the $X$-$s$ orbital on all of the counterions we used $1.402$.

\begin{figure}[t]
\begin{center}
\includegraphics[width=0.7\columnwidth,angle=0]{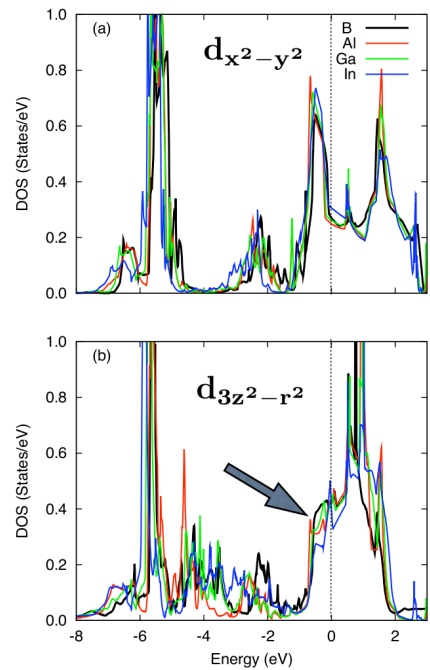}
\caption{$d$-projected density of states (upper panel $x^2-y^2$; lower panel $3z^2-r^2$) of $(001)$ $La_2NiXO_6$ heterostructures with  counter-layer $B$-site ion $X$=$B$ (black) , $Al$ (red), $Ga$  (green) and $In$ (blue), obtained from GGA band theory calculations for unrelaxed structure (ideal cubic perovskite, doubled in $(001)$ direction) with all $Ni-O$ and $X-O$ bond lengths taken to be one half of the bulk $LaNiO_3$ lattice parameter $a=3.81\AA$. Arrow in lower panel indicates region where dependence of $d_{3z^2-r^2}$ density of states on counterion is evident.  }
\label{dosfig}
\end{center}
\end{figure}

\begin{figure}[t]
\begin{center}
\includegraphics[width=0.6\columnwidth, angle=0]{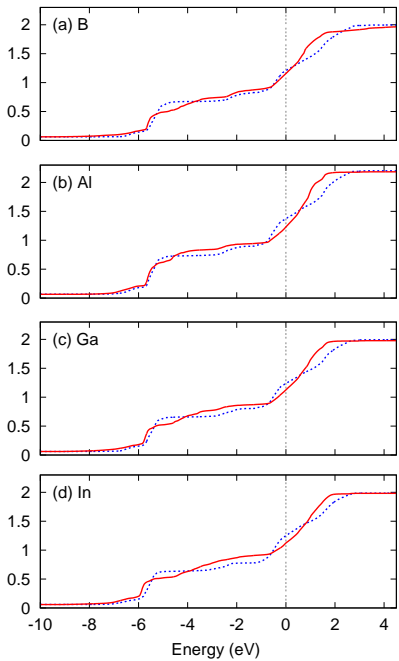}
\caption{Integral of $Ni$-$d_{3z^2-r^2}$ (red solid line) and $Ni$-$d_{x^2-y^2}$ (blue dashed line) density of states for the four  choices of $X$ ion considered in the paper. Flat region between $\sim-2$ and $\sim -1$ $eV$ defines lower edge of conduction band. }
\label{TOS-example}
\end{center}
\end{figure}

Examples of our results for $d$-projected densities of states are shown in Fig. \ref{dosfig}. The density of states consists of two components: a broad antibonding band (of mixed $Ni$-$d$/$O$-$p$ character) spanning the region from $\sim-1.5$ to $\sim 2eV$ near the chemical potential and a narrow bonding band at the low energy $E\sim -6eV$.  As can clearly be seen, the lower edge of the $Ni-O$ antibonding band is well defined. Orbital occupancies are then obtained from integrals of the densities of states; representative results are shown in Fig. \ref{TOS-example}. That the integrals are not quite $2$ is related to the sphere size.  We are interested in the occupancies $n_{x^2-y^2}$ and $n_{3z^2-r^2}$ of the near fermi level $d-p$ antibonding bands. To obtain these we take the difference
between the value of the density of states integral at the lower edge of the antibonding band (identified from the flat part of the integrated density of states plot) and the value at the fermi level. Values corresponding to integrating over the entire conduction band plots of integrated $d$ spectral weight vs energy may be read directly from Fig \ref{TOS-example} and yield the same conclusions.

From the orbital occupancies we obtain the orbital polarization $P$ which we define as
\begin{equation}
P=\frac{n_{x^2-y^2}-n_{3z^2-r^2}}{n_{x^2-y^2}+n_{3z^2-r^2}}
\label{Pdef}
\end{equation}

\section{Results: unrelaxed structure \label{Results}}

Table \ref{Ptable} presents our computed results for the orbital polarization. In this section we focus on the second column, giving results for  `unrelaxed' structures in which all $Ni-O$ and $X-O$ bonds set equal to one half of the $Ni-Ni$ distance of bulk $LaNiO_3$. Results from these structures highlight the chemical effect of interest here. Results for  `z-relaxed' structures obtained by minimizing the energy with respect to atomic motions along the $(001)$ (superlattice) direction and for structures with additional in-plane strain are discussed in section \ref{Relax}.

\begin{table}[htdp]
\begin{center}
\begin{tabular}{c|c|clcl|clclclcll}
\hline
&X&$P_{unrelaxed}^{LNO}$&$P_{relaxed}^{LNO}$&$P_{relaxed}^{STO}$& \\
\hline
&B&.15&-.19&-.10&\\
&Al&.25&.30&.40&\\
&Ga&.26&.33&.42&\\
&In&.36&.41&.57&\\
\hline
\end{tabular}
\end{center}
\caption{Orbital polarization calculated as defined in Eq (\ref{Pdef})  for three superlattice families: `unrelaxed LNO', with all $Ni-O$  and $X-O$ bond-lengths set equal to the GGA-optimized pseudocubic $LaNiO_3$  value of $1.905\AA$ ; `z-relaxed LNO', with in-plane $Ni-O$ and $X-O$  bonds set equal to $1.905\AA$ and out of plane bonds relaxed to  minimize the GGA band theory energy and `relaxed STO' in which the  in-plane $Ni-O$ and $X-O$ bonds are set equal to $1.95\AA$ and out  of plane bond lengths are relaxed to minimize the GGA band theory  energy.  }
\label{Ptable}
\end{table}%

The calculated polarizations are seen to vary strongly with choice of counterion $X$.  The polarization differences can also be seen directly in Fig. \ref{dosfig}, for example as an X-dependent change in the $d_{3z^2-r^2}$ density of states in the lower portion of the antibonding band (indicated by arrow). We stress that the polarization differences occur even though all $Ni-O$ and $X-O$ bond lengths are equal; thus the difference is a chemical effect.  We also stress that while the magnitude of the polarization depends on calculational details such as the sphere sizes and the range over which one integrates, the trends between materials are robust and clearly demonstrate that a non-structural difference between the different $X$ ions strongly influences the orbital polarization.

\begin{figure}[t]
\begin{center}
\includegraphics[width=0.7\columnwidth,angle=0]{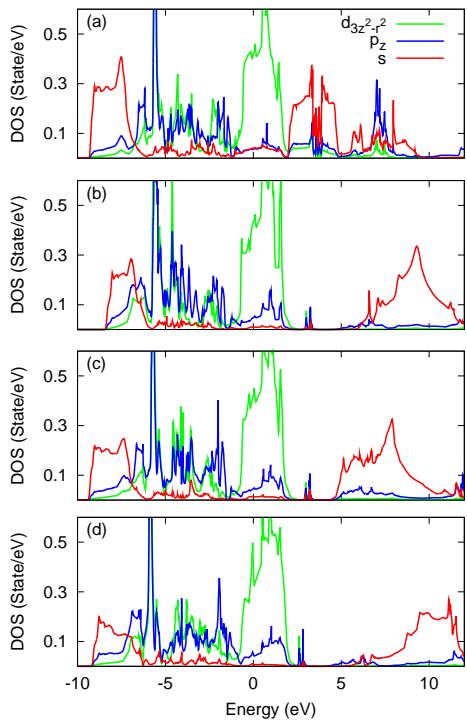}
\caption{Orbital symmetry-projected densities of states for unrelaxed $La_2NiXO_6$ heterostructures with $X$=$B$ (top panel), $Al$ (second from top), $Ga$ (third from top) and $In$ (bottom). Shown are the  $3z^2-r^2$ orbital on the $Ni$ (green trace), the $p_z$ orbital on the apical oxygen (blue trace) and the $s$ orbital on the $X$ site (red trace). 
}
\label{sdosfig}
\end{center}
\end{figure}

We believe that this non-structural difference is related to the properties of a near-fermi-level orbital on the $X$ site. To investigate this hypothesis we computed orbitally projected densities of states for the different systems. We found the most significant effect comes from the $s$-symmetry orbital on the counterlayer $X$ site. In Fig. \ref{sdosfig} we present the $s$ projected density of states on the $X$ site, along with the $Ni$ $3z^2-r^2$ and apical oxygen $p_z$ projected densities of states.  Examination of the series $B$, $Ga$,  $In$ indicates a clear correlation between the orbital polarization and the energy of the $X$-site $s$-symmetry orbital and its hybridization with the apical oxygen orbital. The $Al$ case is an outlier in this series, for reasons not yet understood.   

Higher orbital polarization ($In$ case) is associated with an $s$ orbital which is farther removed in energy and less strongly admixed with $O_{pz}$. Higher orbital polarization is also associated with stronger mixing between the $Ni$ $3z^2-r^2$ orbital and the apical $O_{p_z}$. Lower orbital polarization ($B$ case) is associated with an $s$ orbital which is closer to the fermi level in energy and more strongly admixed with $O_{p_z}$ and is also associated  with weaker mixing between the $d_{3z^2-r^2}$ and the apical $O$. 

Our finding of a key role played by the $s$ orbital on the $X$ site is reminiscent of  results of O. K. Andersen and collaborators,\cite{Pavarini01}. These authors  argued that in the  high $T_c$ cuprate case the variation of the fermiology across different sub-families of cuprate materials was controlled by the energy of the  $Cu$ $4s$ orbital, which affected the ratio of first and second nearest neighbor hopping $t{'}/t$ (similar arguments relating the fermiology of orbitally polarized nickelate heterostructures to the $Ni$ $3z^2-r^2$ orbital were made by Chaloupka and Khalliulin \cite{Chaloupka08}). Here we argue in an analogous way that the variation of the energy of the $X$-site $s$ orbital  controls the polarization of the $Ni$ d-orbitals. (We also investigated the  $Ni$ $4s$ orbital, finding that its energy  does not vary significantly across the series).

We suggest that the polarization differences are due ultimately to changes in the hybridization between the transition metal ion and the nearby oxygen orbitals. In the $La_2NiXO_6$ systems the crucial role is played by the apical oxygen which as can be seen from Fig. \ref{geometry} connects the $s$ and $Ni$ and couples to the $Ni$ $3z^2-r^2$ orbital but not the $x^2-y^2$ orbital.  Fig. \ref{sdosfig} shows that the different choices for ion $X$ have states with differing overlap with the $p_z$ state on the apical oxygen. Thus in essence the counterion affects the polarization by shifting the properties of the apical oxygen $p_z$ state.  

\section{Tight Binding Analysis \label{tb}}

\begin{figure}[b]
\begin{center}
\includegraphics[width=0.7\columnwidth,angle=0]{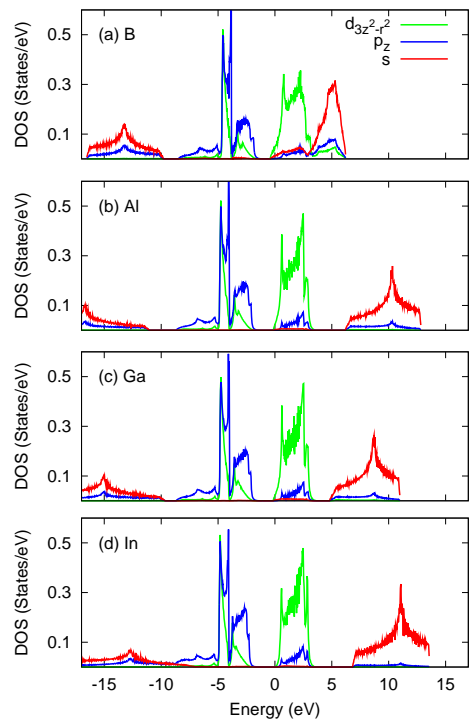}
\caption{ Density of states computed from tight binding model projected onto $Ni$ $d_{3z^2-r^2}$, apical $O_{p_z}$ and $X_s$ states.
}
\label{tbdosfig}
\end{center}
\end{figure}

To confirm the hypothesis that the trends reported in Fig. \ref{sdosfig} are causally related to the observed polarization changes we turn to a tight binding model which, while necessarily a simplified description of the actual band structure, captures with reasonable fidelity the essential features of the density of states. The specifics of the tight binding model are given in the Appendix. We present here the main ideas. We begin with a $5$-band model of cubic $LaNiO_3$ consisting of the two $e_g$ symmetry $Ni$ $d$-orbitals and the three oxygen $p_\sigma$ orbitals.  We find that a very good representation of the GGA-calculated $Ni$ $e_g$ and $O_{p\sigma}$ density of states for cubic $LaNiO_3$ is obtained with a $Ni$ $d$-level energy of $-1.22eV$, an oxygen energy of $-5.2eV$, and
$Ni-O$ and $O-O$ hopping amplitudes of $1.8 eV$ and $0.7eV$ respectively. We next double the unit cell in the $z$ direction and replace one of the two $Ni$ with an $X$ orbital.  We must then introduce three new tight binding parameters: an energy $\varepsilon_X$ of the orbital on the $X$ site and hopping parameters $t_{SP}, t_{SPZ}$ coupling the X orbital to the oxygen ions in the $X$ plane and to the apical oxygens. The parameters $\varepsilon_X$, $t_{SP}$, and $t_{SPZ}$ are numerically optimized by minimizing the difference between the GGA and tight binding (TB) densities of states of  $Ni_{3z^2-r^2}$, apical $O_{pz}$ and $X_S$  in the near fermi level  and positive energy regions. Results are shown in Fig. \ref{tbdosfig}. We see that the tight binding model reproduces the behavior of the near fermi level states well.  Differences (for example an overestimate of the width of the low energy $E\sim-12eV$ portion of the $s$ band and an underestimate of the width of the high energy $E\sim 8eV$ portion) are present, but total number of $X$-$s$ states in the lower and upper energy portions agree well with the GGA calculations (see Appendix).

The orbital polarizations were then computed from the tight binding model, with the results $P_{B}=0.19$, $P_{Al}=0.28$, $P_{Ga}=0.31$ and $P_{In}=0.39$. Comparison to the information presented in Table \ref{Ptable} shows that the tight binding polarizations reproduce well the trends found in the GGA calculation. Because the only `moving part' in the tight binding model is the  $X$ orbital  the successful fitting confirms that the variation in orbital polarization between materials is in fact due to changes in nominally unfilled orbitals on the $X$ site.

\section{Effect of structural relaxations \label{Relax}}

We finally consider the interplay of structural relaxations with the effects we have studied so far in this paper. We performed a structural relaxation process in which atoms were allowed to move in the direction transverse to the plane so as to minimize the $GGA$ energy with the in-plane lattice constants kept equal to the bulk $LaNiO_3$ value.  Table \ref{structure} presents the  $Ni$-apical $O$ and $X$-apical $O$ bond lengths found after relaxation. Note that the $Ni$-apical $O$ bond length is remarkably robust: almost all of the bond length change due to relaxation occurs in the $X$-$O$ bond. We take this as further confirmation of the `chemical effect': the geometrical $Ni-O$ distance is less important than what happens at the other end of the $Ni-O-X$ bond.  The corresponding polarizations are given in the third column of Table \ref{Ptable}. Relaxation has a large effect on $P$, and acts to enhance the `chemical' effects we have identified. Note in particular that the sign of the polarization actually reverses in the $B$ case.

\begin{figure}[t]
\begin{center}
\includegraphics[width=0.7\columnwidth,angle=0]{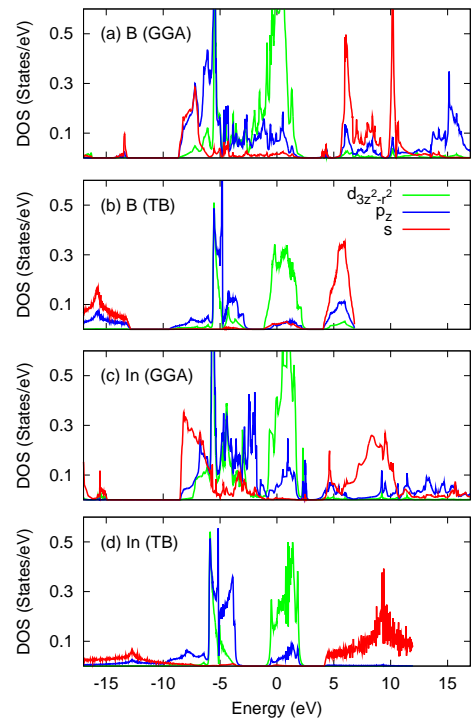}
\caption{GGA orbital symmetry-projected densities of states for relaxed $La_2NiXO_6$ heterostructures with $X$=$B$ (top panel),  and $In$ (third from top), along with tight binding fits for the two cases (second and fourth panels from the top). Shown are the  $3z^2-r^2$ orbital on the $Ni$ (green trace), the $p_z$ orbital on the apical oxygen (blue trace) and the $s$ orbital on the $X$ site (red trace). 
}
\label{relaxdosfig}
\end{center}
\end{figure}

The first and third panels of Fig \ref{relaxdosfig} show the GGA densities of states of the two extreme cases ($B$ and $In$)  computed for the relaxed structure. Comparison to Fig. \ref{sdosfig} shows  that in the boron case the $X_S$ density of states shifts upwards in energy, while in the In case the $X_S$ density of states shifts downwards in energy, as expected if the position of the $X$ orbital is controlled by hybridization with the oxygen $p_z$.

\begin{table}[htdp]
\begin{center}
\begin{tabular}{|c|c||cc|lcc|l}
\hline
&X&$d^{Ni-O}_{LNO}$&$d^{X-O}_{LNO}$&$d^{Ni-O}_{STO}$&$d^{X-O}_{STO}$& \\
\hline
&B&1.07&0.85&1.05&0.80&\\
&Al&1.00&1.00&0.99&0.98&\\
&Ga&1.00&1.06&0.99&1.05&\\
&In&0.99&1.19&0.97&1.16&\\
\hline
\end{tabular}
\end{center}
\caption{Bond lengths between $Ni$ and apical oxygen and between
  $X$-site atom and apical oxygen presented as ratio of computed bond
  length to $LaNiO_3$ value $a=3.81\AA$ for relaxed structures with
  in-plane lattice constant set to $LaNiO_3$ value ($LNO$) and to
  $SrTiO_3$ value $a=3.905$ ($STO)$. }
  \label{structure}
\end{table}%

To show how the effects of structural relaxations are incorporated
within the tight binding model we have used the same TB parameters as
in the unrelaxed case, except that we have increased $t_{SPZ}$ in the
B case (modelling the effect of a decreased $X-O$ distance and
decreased it in the $In$ case (modelling the effect of an increased
$X-O$ distance.  The resulting tight binding densities of states are
also shown in Fig. \ref{relaxdosfig}. The tight binding model
reproduces the basic shifts in the $X$ orbital density of states and
leads to polarizations, $P^{TB}_{B}=0.08$ and $P^{TB}_{In}=0.44$.

Finally, to study the effects of in-plane strain (induced, for example by growing on a different substrate) we set the in-plane lattice constant equal to the value $3.91\AA$ appropriate to the widely-used $SrTiO_3$ substrate material. This corresponds to applying a tensile in-plane strain.  Results for the polarization are shown in the fourth column of Table \ref{Ptable} and for the lattice constants in the fourth and fifth columns of Table \ref{structure}. The tensile strain affects the polarization in the expected way but the out of plane lattice constant is again remarkably robust.

\section{Conclusion \label{Conclusion}}
Thus in summary we have shown that in addition to the well-established geometrical effect, orbital polarization in an oxide heterostructure may be controlled by appropriate choice of counterions in the superlattice. Changes in the counter-ion produced large orbital polarization differences, which are expected to be further enhanced by correlation effects (not yet included in our calculations).  Both the magnitude and (after relaxation) the sign of the polarization can be changed. The crucial factor was found to be the hybridization of an $s$-like orbital on the counter-ion with the $p_z$ orbital on the apical oxygen.  This hybridization changed the interaction of the apical oxygen $p_z$ with the transition metal ion, and hence changed the orbital polarization.  We also found that under structural relaxation the $Ni-O$ bond lengths change relatively little, while the $X-O$ bonds changed substantially. Our results thus show that the transition-metal/oxygen bond length is not the only variable controlling polarization, and that chemical effects should in general be considered when attempting to optimize superlattice properties. We remark that the need to include the $s$ orbital demonstrates a limitation of the common theoretical strategy of deriving from band
structure a `minimal low energy model' of the system of interest. While of course a careful downfolding, keeping track of the effects stemming from bands that are projected out, will produce a low energy model which contains the effects of interest (see, e.g. \cite{Pavarini01,Andersen95}), construction of `minimal high energy models' such as the one we have defined may also be a useful strategy.

\begin{acknowledgments}
This work was supported by the U. S. Army Research Office via grant No. W911NF0910345 56032PH. Part of the work reported here was performed while one of us (AJM) was in residence at the KITP with part support from the National Science Foundation under Grant No. PHY05-51164.
\end{acknowledgments}

\vspace{.5in}

{\bf Appendix: Details of tight binding model}

\vspace{.2in}

The tight binding model presented in the text is based on the minimal assumption of a cubic $ABO_3$ perovskite structure, with the unit cell doubled in the $(001)$ direction and the two $B$-sites distinguished. It involves 9 orbitals: $Ni$ $x^2-y^2$ and $3z^2-r^2$, the $O_{p\sigma}$ orbitals in the $Ni$-$O$ plane ($O_{Ni,x}$, $O_{Ni,y}$), $O_{p\sigma}$ orbitals ($O_{ap1,z}$, $O_{ap2,z}$) at the two apical oxygen sites, an $s$ orbital and two $O_{p\sigma}$ orbitals in the $X-O$ plane ($O_{X,x}$, $O_{X,y}$).

Ordering the basis states as $(3z^2-r^2,x^2-y^2,O_{Ni,x},O_{Ni,y},O_{AP1,z},O_{X,x},O_{X,y},O_{AP2,z},X)$ we may write the tight binding Hamiltonian $H_{TB}$ in a schematic block-diagonal form as
\begin{equation}
H_{TB}=\left(\begin{array}{ccc}H_{Ni} & H_{PD}^\dagger & 0 \\H_{PD} &
  H_O & H_{PX}^\dagger \\0 & H_{PX}& H_{X}\end{array}\right)
\label{Htb}
\end{equation}

We took $H_{Ni}=\varepsilon_d {\mathbf I}_2$with $I$ the $2\times2$ unit matrix and $\varepsilon_d=-1.22eV$.

For the oxygen portion $H_O$ we assumed the usual overlaps between orbitals on second neighbor oxygens, wrote the form appropriate for a simple cubic lattice, and then doubled the unit cell along $(001)$. To avoid a cumbersome display of $6\times 6$ matrices we write the result before doubling as

\begin{widetext}

\begin{equation}
H_p=\varepsilon_P{\mathbf I}_3-t_{PP}\left(\begin{array}{ccc}
0 &\left(1-e^{ik_x}\right) \left(1-e^{-ik_y}\right)&\left(1-e^{ik_x}\right) \left(1-e^{-ik_z}\right)\\
\left(1-e^{-ik_x}\right) \left(1-e^{ik_y}\right)& 0 &\left(1-e^{ik_y}\right) \left(1-e^{-ik_z}\right)\\
\left(1-e^{-ik_x}\right) \left(1-e^{ik_z}\right) & \left(1-e^{ik_z}\right) \left(1-e^{-ik_y}\right)& 0
\end{array}\right)
\end{equation}

\end{widetext}

Here ${\mathbf I}_3$ is the $3\times 3$ unit matrix and we took $\varepsilon_P=-5.2 eV$ and $t_{PP}=0.7eV$. Inclusion of oxygen-oxygen hopping is necessary to reproduce the narrow $d$ feature observed at $E\sim -6eV$ in e.g. Fig. 2 of the main text. $H_O$ is just $H_P$ doubled in the $z$ direction.

For the $p-d$ hopping we took the standard form
\begin{equation}
H_{pd}=-t_{pd}
\left(\begin{array}{cc} \frac{1}{2}\left(1-e^{ik_x}\right)  &- \frac{\sqrt{3}}{2}\left(1-e^{ik_x}\right) \\
 \frac{1}{2}\left(1-e^{ik_y}\right)  &  \frac{\sqrt{3}}{2}\left(1-e^{ik_y}\right)  \\
 -e^{-i\frac{k_z}{2}} & 0 \\
 0 & 0 \\
0 & 0 \\
  e^{i\frac{k_z}{2} } & 0\end{array}\right)\label{Htpd}
\end{equation}
with $t_{PD}=1.8eV$ while for $H_{XP}$ we put
\begin{equation}
H_{PX}=
\left(\begin{array}{c} 
0  \\
 0  \\
  -t_{SPZ} e^{i\frac{k_z}{2} }\\
 t_{SP}\left(1-e^{-ik_x}\right)  \\
t_{SP}\left(1-e^{-ik_y}\right)   \\
t_{SPZ}  e^{-i\frac{k_z}{2}}
\end{array}\right)\label{HSP}
\end{equation}

The important parameter is $t_{SPZ}$ giving the overlap between the $s$ orbital and the apical oxygen states.

The tight binding parameters are not uniquely determined; changes in
one parameter may be to some extent compensated by changes in another,
but as long as the density of states and $s$-apical $p_z$ mixing are
reproduced with reasonable accuracy the polarization is robust. We
chose to fix the tight binding model parameters by first choosing the
$Ni$-related parameters to fit the GGA band structure of cubic
$LaNiO_3$. Then we determined the other orbitals by minimizing the
'distance' between the GGA and TB predictions for the density of
states of the $3z^2-r^2$, $Op_z$ and $X_S$ orbitals. That is, we chose
the TB parameter set which minimizes $\sum_{\mathcal O}\int d\epsilon
(\rho(\epsilon)^{{\mathcal O}}_{GGA}-\rho(\epsilon)^{{\mathcal
    O}}_{TB})^2$ where $\rho(\epsilon)^{\mathcal O}$ denotes the
density of states for orbital ${\mathcal O}=Ni ~d_{3z^2-r^2}, ~O_{pz}$
or $X_S$ atom and the energy ranges are $-1$ to $2$ for the $Ni$,
$-10$ to the upper-limit for the $O_{pz}$ and $-2$ to the upper limit
$X_S$ orbitals.  The parameters used to produce the fits shown in the
text for the unrelaxed structures are summarized in Table \ref{TBpol}
along with the resulting polarizations. To model the effects of
structural relaxation in the $In$ case we decreased the $t_{SPZ}$ from
$4$ to $2.5$ and in the $B$ case we increased it from $4.5$ to
$6$. The resulting polarizations are $P^{TB}_{B}=0.08$ and
$P^{TB}_{In}=0.44$.

\begin{table}[htdp]
\begin{center}
\begin{tabular}{|c|ccc|c|lcc|l}
\hline
 & $t_{SP}$ & $t_{SPZ}$ & $\varepsilon_X$ &P& \\
 \hline
 B & 3.0 & 4.5 & -4.0 & 0.19 \\
Al & 5.0 & 6.0 & -1.6 & 0.28 \\
Ga & 4.5 & 5.0 & -1.6 & 0.31  \\
In & 4.5 & 4.0 &  3.2 & 0.39 \\
\hline
\end{tabular}
\end{center}
\caption{Tight binding parameters used to construct fits displayed in
  Figs. 4 of main text, with resulting orbital polarization }
  \label{TBpol}
\end{table}

\begin{figure*}[t]
\begin{center}
\includegraphics[width=0.7\columnwidth, angle=0]{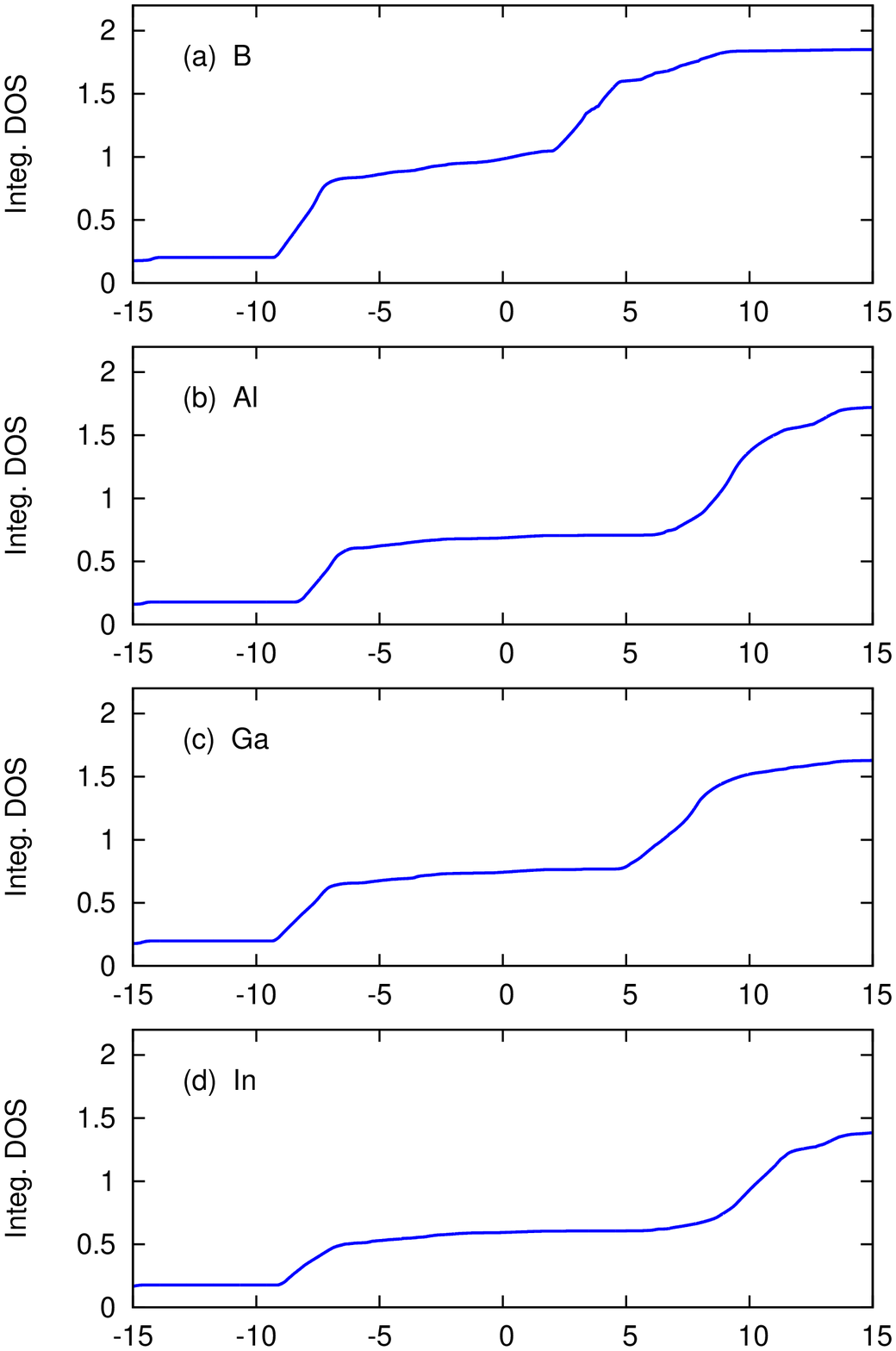}
\includegraphics[width=0.7\columnwidth, angle=0]{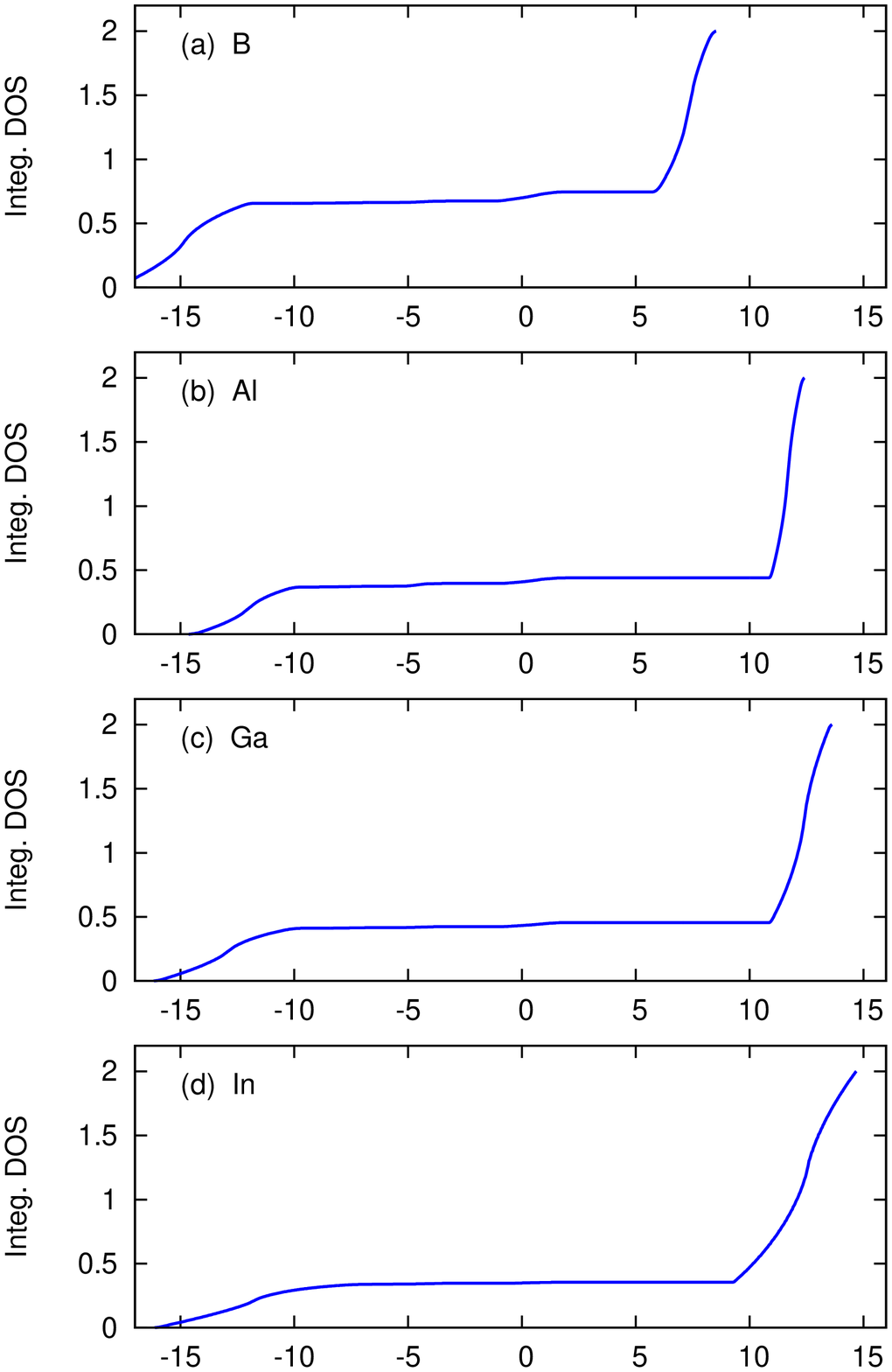}
\caption{Left panel: integral of $X$-$s$ GGA density of states for the four choices of
  $X$ ion considered in the paper. Right panel: integral of $X$-$s$ TB density of states for the four choices
  of $X$ ion considered in the paper}
\label{TB1}
\end{center}
\end{figure*}

Figure \ref{TB1} compares the integrated $X$-$s$ density of states
obtained from GGA and optimized TB calculations respectively. Some
differences are visible. At the very low end of the energy range ( $E
\sim -8$ to $-12 eV$ the $X$ density of states extends to low energies
in the tight binding model than it does in the GGA calculation,
presumably because of level repulsion arising from other low energy
orbitals present in the GGA calculations but not in the tight binding
model. Similarly, at the very high end of the range the tight binding
model underestimates the width of the $S$ band, but these differences
do not affect the qualitative conclusions we wish to draw.

\vspace{2.in}

\bibliography{refs_orbitalcontrol1.bib}

\end{document}